%% file: main_arxiv.tex
\documentclass[10pt,letterpaper,conference,nofonttune]{IEEEtran}
\IEEEoverridecommandlockouts

\usepackage{cite}
\usepackage{amsmath,amssymb,amsfonts}
\usepackage{algorithmic}
\usepackage{graphicx}
\usepackage{textcomp}
\usepackage{listings}
\usepackage{soul}
\usepackage{url} 
\usepackage{listings}
\usepackage{xcolor}
\usepackage{etoolbox}
\usepackage{tikz}
\usetikzlibrary{calc}
\usepackage{xspace}
\usepackage{tabularx}  
\usepackage{booktabs}  
\usepackage{array}     
\usepackage{ragged2e}
\usepackage{subcaption}
\usepackage{multirow}  
\usepackage[draft]{hyperref}
\usepackage{balance}

\usepackage[acronyms,nonumberlist,nopostdot,nomain,nogroupskip,acronymlists={hidden}]{glossaries}
\newglossary[algh]{hidden}{acrh}{acnh}{Hidden Acronyms}
\glsdisablehyper
\input{acronyms.tex}

\lstdefinestyle{mystyle-json}{
  commentstyle=\color{black},
  keywordstyle=\color{black},       
  numberstyle=\tiny\color{blue},
  stringstyle=\color{black},        
  basicstyle=\ttfamily\scriptsize,
  rulecolor=\color{black},
  breakatwhitespace=true,         
  breaklines=true,                 
  captionpos=b,
  frame=tb,
  keepspaces=true,                 
  numbers=left,                    
  numbersep=5pt,                  
  showspaces=false,                
  showstringspaces=false,
  showtabs=false,                  
  tabsize=2,
  xleftmargin=10pt,
  belowskip=-10pt,
}

\lstdefinelanguage{json}{
  alsoletter={-},
  keywords={listen,on,off,dst,periodic},
  sensitive=false,
  comment=[l]{\#},
  morecomment=[s]{/*}{*/},
  moredelim=**[il][\color{black}{:}]{:}
  morestring=[b]',
  morestring=[b]",
}

\def\BibTeX{{\rm B\kern-.05em{\sc i\kern-.025em b}\kern-.08em
    T\kern-.1667em\lower.7ex\hbox{E}\kern-.125emX}}

\def\framework{TENORAN\xspace}

\begin{document}

\title{TENORAN: Automating Fine-grained Energy Efficiency Profiling in Open RAN Systems}

\author{\IEEEauthorblockN{
Ravis Shirkhani\IEEEauthorrefmark{1},
Stefano Maxenti\IEEEauthorrefmark{1},
Leonardo Bonati\IEEEauthorrefmark{1},
Niloofar Mohamadi\IEEEauthorrefmark{1},
Maxime Elkael\IEEEauthorrefmark{1},\\
Umair Hashmi\IEEEauthorrefmark{2},
Jeebak Mitra\IEEEauthorrefmark{2},
Michele Polese\IEEEauthorrefmark{1},
Tommaso Melodia\IEEEauthorrefmark{1},
Salvatore D'Oro\IEEEauthorrefmark{1}
}
\IEEEauthorblockA{
\IEEEauthorrefmark{1}
Institute for the Wireless Internet of Things, Northeastern University, Boston, MA, U.S.A.\\
Email: \{shirkhani.r, maxenti.s, l.bonati, n.mohamadi, m.elkael, m.polese, t.melodia, s.doro\}@northeastern.edu\\
\IEEEauthorrefmark{2}Dell Technologies Inc.
\hspace{5pt}
Email: \{umair.hashmi, jeebak.mitra\}@dell.com}
\thanks{This work was partially supported by the National Telecommunications and Information Administration (NTIA)'s Public Wireless Supply Chain Innovation Fund (PWSCIF) under Award No. 25-60-IF002.}
}

\maketitle

\ifnumequal{\thepage}{1}{%
    \begin{tikzpicture}[remember picture,overlay]
        \node[draw,
        minimum width=0.5\paperwidth,
        text width=0.5\paperwidth,
        align=center,
        font=\scriptsize,
        anchor=north
        ]
        at ($(current page.north)+(0,-15pt)$)
        {%
        This article has been accepted for publication in IEEE INFOCOM WKSHPS: NG-OPERA 2026: 
        Next-generation Open and Programmable Radio Access Networks. This is the author's version 
        which has not been fully edited and content may change prior to final publication.
        };
    \end{tikzpicture}%
}{}

\begin{abstract}
The transition to disaggregated and interoperable Open \gls{ran} architectures and the introduction of \glspl{ric} in O-RAN creates new resource optimization opportunities and fine-grained tuning and configuration of network components to save energy while fulfilling service demand. However, unlocking this potential requires fine-grained and accurate energy measurements across heterogeneous deployments. Three factors make this particularly challenging. First, data collection and energy profiling require extensive, repeatable end-to-end tests that must coordinate heterogeneous components and automatically collect power and performance measurements to evaluate energy efficiency. Second, softwarization in O-RAN enables continuous updates and improvements, but results in frequent code releases that might alter energy consumption profiles and calls for continuous data collection and testing. Third, no single power measurement tool can observe all parts of an O-RAN deployment that spans hardware and software domains: rack-level power distribution units, container-level estimators, and dedicated power meters for radio units field-deployed at cell sites all provide partial and heterogeneous views that are difficult to integrate into a unified measurement pipeline.

To address these challenges, we design the \framework framework, an automated measurement scaffold for fine-grained energy efficiency profiling of O-RAN deployments, and prototype it on a heterogeneous OpenShift cluster. \framework instruments an end-to-end deployment based on high-level specifications (e.g., gNB software stack and split options, traffic profiles), and collects synchronized performance metrics and power measurements for individual \gls{ran} components while the network is under controlled workloads including over-the-air traffic. Our experimental results demonstrate energy profiling of end-to-end experiments with xApps in the loop, energy efficiency differences between two \gls{ran} stacks, \acrlong{oai} and srsRAN, in uplink and downlink, and core network power consumption trends.
\end{abstract}


\glsresetall
\glsunset{usrp}
\glsunset{json}

\begin{figure*}[t!]
\setlength\abovecaptionskip{1pt}
    \centering
    \includegraphics[width=0.9\textwidth,keepaspectratio]{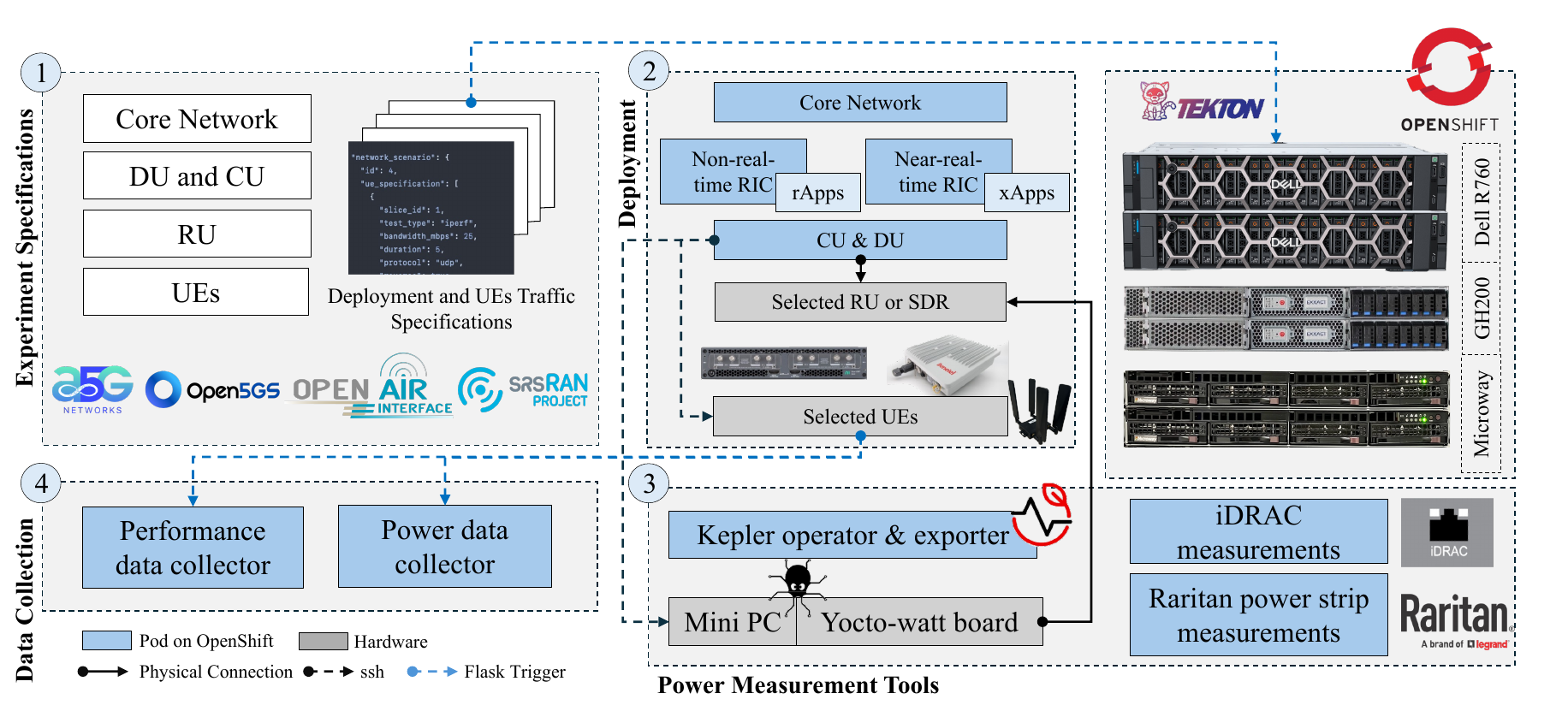}
    \caption{High-level overview of \framework pipeline to automatically collect performance and power measurements.}
    \label{fig:pipeline}
\end{figure*}

\section{Introduction}
\label{sec:introduction}

A recent study from Ericsson~\cite{ericsson_mobility_2025} shows that the \gls{ict} sector as a whole is estimated to consume around 4\% of global electricity, corresponding to nearly 1{,}100~TWh in 2024. The same data also shows that networks play a major role with more than 300~TWh consumed in 2024.
In parallel, GSMA estimates that mobile network operators consumed around 290 TWh of electricity in 2023 (about 1\% of global electricity use).
The carbon emissions related to \gls{ran} operations fell by about 8\% between 2019 and 2023, despite a 9\% increase in new 5G users, and mobile data traffic nearly quadrupling over the same period~\cite{gsma_mobilenetzero_2025} which reflects the adoption of renewable electricity procurement and network energy-efficiency measures. However, the continued growth in traffic and infrastructure footprint means that further efficiency gains remain necessary.
In this context, it is important to mention that the \gls{ran} accounts for over three-quarters of the total energy consumption of mobile networks~\cite{GSMA_GoingGreen}, and its cost constitutes 25\% of operators' \gls{opex}~\cite{GSMA_VPP_2024}. For this reason, improving \gls{ran} energy efficiency is both an environmental and an economic priority.

At the same time, the need for more agile, efficient and easy-to-maintain, operate and upgrade cellular networks has motivated a gradual transition from traditional to Open \gls{ran} deployments. The Open \gls{ran} architecture introduces unprecedented flexibility and innovation~\cite{10024837} via multi-vendor, disaggregated components. 
Since this approach is rooted in virtualization, cloud-native principles and increasingly more powerful hardware accelerators~\cite{kelkar_nvidia_2021}, multiple \gls{ran} workloads can be hosted on the same shared infrastructure~\cite{bonati2023neutran}, which promises more efficient resource utilization and increased energy savings via dynamic and demand-aware resource allocation and infrastructure scaling~\cite{kundu2024energyefficientranindustry}. 

Indeed, to achieve sustainability and optimized energy management, a major requirement is that of being able to measure energy consumption of each network element at high-resolution.
However, due to softwarization and infrastructure sharing, and architectural disaggregation, traditional server-level, wrap-around power measurement tools are no longer sufficient, as they fail to capture statistics of individual virtualized components, e.g., \glspl{cu} and \glspl{du}, which may be deployed with different multiplicities on the same server, or distributed across servers in one or multiple clusters. For this reason, how to measure the energy consumption of these new cloud-native and disaggregated architectures, such as those promoted by Open \gls{ran}, with frequent software updates that might affect their energy consumption is still an open problem, which makes it hard to truly assess energy savings and demonstrate their sustainability claims~\cite{ahmadi_toward_2025}. This complexity is further exacerbated by the growing adoption the O-RAN \glspl{ric} and their intelligent applications (e.g., xApps and rApps)~\cite{maxenti_scalo-ran_2024}. Indeed, these can set policies to reduce \gls{ran} energy consumption~\cite{lozano_kairos_2025}, while simultaneously introducing software components whose energy needs to be accounted for and measured. Moreover, in disaggregated \gls{ran} architectures, \glspl{ru} are often not co-located with the \gls{cu}/\gls{du} and may be deployed at remote cell sites or edge locations, operating under distinct power domains and measurement constraints. As a result, \gls{ru} energy consumption cannot be inferred from data-center measurements and instead requires dedicated, site-level power monitoring mechanisms.

The increasing complexity of O-RAN deployments and the need for building sustainable and energy-efficient cellular networks calls for integrated, automated frameworks that can facilitate energy profiling of the different components of an \gls{e2e} O-RAN system under varying loads and conditions. 
Designing such frameworks is non-trivial, as it requires coordinating heterogeneous measurement sources across disaggregated hardware and virtualized software components, ensuring temporal alignment between power and performance metrics, and maintaining reproducibility across dynamic, continuously evolving deployments.
Such frameworks must also therefore include effective data-collection pipelines and monitoring capabilities to capture performance metrics alongside power consumption for individual virtualized workloads and network functions enabling the exploration of the energy–performance trade-off.

\subsection{Contributions}
\label{sec:contributions}
In this paper, we present \framework, a framework for automatically profiling performance and energy of \gls{e2e} O-RAN deployments, which addresses three main challenges. First, the softwarization of O-RAN leads to frequent updates and changes in the energy profile of individual components, which necessitates continuous, automated profiling rather than ad-hoc measurements. Second, power measurements in realistic O-RAN deployments need to come from diverse tools (rack-level \glspl{pdu}, server telemetry, container-level estimators, and dedicated power meters for \glspl{ru}), which need to be integrated into a unified framework that can also show how energy consumption ties with network performance and load. Third, manually deploying heterogeneous O-RAN components, characterizing traffic, and collecting both performance and power metrics is time-consuming and difficult to repeat across scenarios. \framework provides a practical solution to these challenges via a unified, automated framework that takes as input high-level test specifications, configures and deploys an \gls{e2e} O-RAN network, executes \gls{ota} tests, and automatically collects application-layer performance metrics together with power measurements for individual components.

We prototype \framework on a Red Hat OpenShift cluster where we integrate a diverse set of open-source \gls{ran} (e.g., \gls{oai} and srsRAN) and core network (e.g., Open5GS and a commercial core network) solutions, as well as Near-RT \gls{ric} and xApps, demonstrating that it can automate energy profiling across heterogeneous deployments and components.
A high-level view of our framework is shown in Figure~\ref{fig:pipeline}. \framework provides energy measurements at different granularities: (i) server-level power monitoring is obtained via smart \gls{pdu} measurements; (ii) power estimation of virtualized workloads (e.g., \glspl{cu}/\glspl{du}, xApps) via Kepler~\cite{amaral_kepler_2023}; and (iii) power consumption of \glspl{ru} via Yocto-Watt power meters~\cite{yoctowatt_manual}.
By integrating these software and hardware components on a single platform, \framework enables repeatable, automated testing and data collection in O-RAN systems.

The remainder of this paper is organized as follows. In Section~\ref{sec:realted-work}, we review solutions related to our work. In Section~\ref{sec:background}, we describe the
\framework architecture. In Section~\ref{sec:architecture}, we detail \framework workflows and \gls{ota} tests orchestration. Finally, in Section~\ref{sec:experiments}, we present our experimental results, while we draw our conclusions in Section~\ref{sec:conclusion}.

\section{Related Work}
\label{sec:realted-work}
How to tackle the challenge of measuring power consumption in O-RAN deployments, has recently gained interest in the literature. The majority of works focus on specific deployment scenarios or measurement tools rather than providing an \gls{e2e} automated framework with power measurements for individual components. In \cite{shankaranarayanan_poet_2024}, authors present a platform that combines \gls{pdu} and Kepler measurements in a testbed that includes both bare-metal and Kubernetes deployments of \gls{oai} with physical \glspl{ru} and \glspl{sdr}. Their work focuses on building an energy measurement infrastructure but do not provide tools to automate experiments and facilitate energy-profiling of O-RAN components across diverse configurations. The framework presented in~\cite{centofanti_energy_2024} proposes the use of a digital power meter to monitor the energy consumption of an \gls{oai}-based 5G gNB with \glspl{sdr} but does not consider standard solutions based on split~7.2 nor O-RAN components such as \glspl{ric} and xApps/rApps. In a complementary direction,~\cite{centofanti_impact_2024} provides a comprehensive evaluation of open-source power measurement tools in containerized cloud environments, comparing software-based and hardware-based approaches across multiple Kubernetes clusters, but without integrating these methods into a \gls{ran}-specific automation pipeline.

Other works analyze how different \gls{ran} architectures and functional splits impact power consumption. For example,~\cite{gudepu_earnest_2024,gudepu_demonstrating_2024} study monolithic, disaggregated \gls{cu}/\gls{du}, and \gls{cups}-based deployments using tools such as s-tui, Scaphandre, and Kepler on Kubernetes. Other work like~\cite{al-tahmeesschi_enhancing_2025} compare power consumption for Split~8 and Split~7.2b under varying traffic loads using \texttt{powerstat} for containerized network functions, which again yields mostly node-level rather than container-level measurements. Overall, these efforts demonstrate the importance of energy measurements and profiling in O-RAN but also show that there is still a lack of integrated, automated frameworks that can also orchestrate \gls{ota} experiments and collect fine-grained power and performance measurements across heterogeneous \gls{ran} components and virtualized workloads. 

A work relevant to \framework is that in~\cite{maxenti_autoran_2025}, where authors present a framework for deployment and configuration of O-RAN systems. However, \cite{maxenti_autoran_2025} focuses on deployment procedures and does not provide tools for multi-stage, granular and hardware/software-based energy measurements, which is instead the focus of \framework.

The goal of this paper is address this gap and to present \framework, a novel framework for \gls{e2e} energy efficiency measurements in O-RAN that we demonstrate on a real-world O-RAN deployment.

\section{\framework Architecture}
\label{sec:background}

\framework architecture is illustrated in Fig. \ref{fig:pipeline}. It provides the following core functionalities to deliver a unified solution for energy testing of O-RAN networks: (i) a virtualization environment based on containers with the ability to deploy and configure O-RAN components; (ii) a set of energy measurement tools able to capture energy consumption of each individual hardware and software components; and (iii) a catalog of O-RAN components to be tested and measured.

To demonstrate the effectiveness of this architecture, \framework has been prototyped on a private 5G testbed based on OpenShift using general-purpose servers and open-source tools. Hardware and software power measurement solutions are integrated into \framework to capture the power consumption of individual O-RAN network elements.  This section provides background on the above core functionalities as well as the hardware and software components we have used in our prototype.

\subsection{Container and Automation Platform}
\label{sec:openshift-cluster}

A major feature of an automation framework for energy testing in O-RAN is providing a streamlined and programmatic approach to automated configuration and deployment of workloads. In \framework, we achieve this by combining a container-based approach where network functions can be deployed as containerized workloads on a general-purpose infrastructure with a set of automation pipelines that can convert high-level test descriptions (see Section~\ref{sec:test-specifications} for more details) to actionable network deployments. 

To demonstrate these core functionalities, in our \framework prototype we leverage Red Hat OpenShift as the container and automation platform.
OpenShift is a container-based virtualization and orchestration framework for deploying and managing applications (e.g., \gls{ran}, core network and \gls{ric} workloads in our case) in the form of atomic microservices.

We also integrate \gls{ci}/\gls{cd} tools, such as Tekton and ArgoCD, to automate the energy measurement of our testing workflows (Section~\ref{sec:architecture}).
Tekton allows us to define and execute automated pipelines in cloud-native systems, and to run ordered sets of tasks on our the virtualized infrastructure~\cite{maxenti_autoran_2025}.
Similarly, ArgoCD offers declarative GitOps routines to synchronize compute node configurations and experiment definitions from version-controlled repositories, which act as a source of truth.
Together, these tools support an automated pipeline that builds containers on physical servers, retrieves test parameters, executes tests on the cluster, and collects performance and power measurements from the different components of the 5G network without manual intervention. 

The infrastructure used in our prototype consists of compute nodes with heterogeneous architectures (i.e., x86 and ARM), as shown in Table~\ref{tab:list-nodes}.
\begin{table}[b]
    \centering
    \caption{Compute node models and architectures.}
    \begin{tabular}{lll}
        \toprule
        Model & CPU & Memory\\
        \midrule
        Dell R760 & Intel Xeon 8462Y+ & 512GB\\
        Microway EPYC & AMD EPYC 7262 & 256GB\\
        Supermicro ARS-111GL-NHR & Grace CPU & 512GB \\
        \bottomrule
    \end{tabular}
    \label{tab:list-nodes}
\end{table}
We run core networks and Near-RT \gls{ric} on Dell R760 nodes, which also run the control plane of the OpenShift cluster.
\gls{ran} workloads, instead, are instantiated on the worker nodes of the cluster. Specifically, \gls{usrp}-based deployments run on Microway nodes, while GPU-accelerated ones on Supermicro nodes, equipped with NVIDIA GH200 GPUs.

\subsection{Energy Measurement Tools}
\label{sec:energy-measurement-tools}

\subsubsection{Container-level measurements}
\label{sec:kepler}

Virtualized O-RAN deployments run multiple network functions (e.g., \glspl{cu}/\glspl{du}, \glspl{upf}, xApps) as container workloads on a shared compute infrastructure. While smart \glspl{pdu} provide accurate node-level power measurements, they cannot attribute consumption to individual pods or services, making it difficult to understand how much energy each softwarized component consumes. To support fine-grained, container-level energy profiling, power estimation tools that can natively operate in containerized environments (e.g., Kubernetes, OpenShift) and expose per-pod power and energy metrics over time are required.
We select Kepler as such tool in \framework. 

Kepler is a \gls{cncf} sandbox project designed to monitor and export power and energy metrics in Kubernetes-based environments~\cite{amaral_kepler_2023,amaral_process-based_2024}.
We deploy Kepler as a cluster operator and integrate it with the native OpenShift monitoring stack as a Prometheus exporter.
Then, we leverage Grafana to visualize power statistics via a dedicated dashboard. This includes metrics related to node architectures, power sources, total energy consumption, and namespaces and pods.
Internally, Kepler combines low-level hardware and kernel metrics with power models to estimate process-level power consumption, which is then aggregated at the container and pod levels~\cite{amaral_kepler_2023}. It leverages \gls{ebpf} routines in the Linux kernel to collect performance counters (e.g., CPU cycles, instructions, cache events) and operating-system metrics, and correlates them with real-time power readings from interfaces such as \gls{rapl} for CPUs and DRAM, \gls{acpi} and Redfish sensors for platform power, and \gls{nvml} for GPUs. Using these inputs, Kepler can derive idle, activation, and dynamic power, and distribute node power across processes and containers~\cite{amaral_process-based_2024}.

\subsubsection{Server-level measurements}
\label{sec:raritan}
Apart from monitoring power consumption of containers and other software components of the O-RAN ecosystem, it is also important to measure the energy consumption of the compute infrastructure. To achieve this, \framework leverages metered \glspl{pdu} distributing power to the infrastructure.

In our OpenShift-based \framework prototype, we leverage Raritan PX4 \glspl{pdu} for this purpose~\cite{raritan_px4}. The PX4 provides outlet-level measurements of voltage, current, active power, and energy with a metering accuracy of $\pm 0.5\%$ and sampling granularity of $1$\:s, and exposes a web interface for monitoring and configuration.
We deploy dedicated collector pods that periodically query the \glspl{pdu} for measurements associated with the outlets feeding each node and store the resulting time series data in an InfluxDB database. Data is then visualized through Grafana dashboards, complementing the container-level metrics captured by Kepler with server-level power measurements.
In addition to these \gls{pdu}-based measurements, we leverage the \gls{idrac} to derive telemetry for the Dell servers of our cluster, as well as their thermal status. Similarly to the \gls{pdu}-level metrics, these are stored in InfluxDB and visualized through Grafana dashboards.

Although these server-level measurement techniques enable accurate, hardware-level power monitoring independent of the software running on the servers, their low time resolution (e.g., $1$\:s for \gls{pdu}-based measurements) make them unsuitable for capturing short-lived power variations (e.g., per-slot load dynamics). This complicates fine-grained energy profiling and motivates the use in \framework of additional, higher-resolution measurement tools, which will be described in Section~\ref{sec:yocto-watt}.

\subsubsection{\gls{ru}-level measurements}
\label{sec:yocto-watt}
\glspl{ru} power consumption can vary at the millisecond scale with transmission patterns such as \gls{prb} allocation, slot utilization, and antenna configuration.
Measuring tools with larger granularities, such as the ones based on metered \glspl{pdu}, cannot, therefore, capture short-lived power variations or the impact of configuration changes.
To obtain such high-resolution traces, \framework integrates Yocto-Watt power meters connected to each \gls{ru}.

The Yocto-Watt is a non-invasive USB-connected wattmeter that measures DC/AC voltage and current and reports instantaneous power and accumulated energy~\cite{yoctowatt_manual}. In our setup, a Yocto-Watt module is inserted in series with the power supply of each \gls{ru}, and is connected via USB to a nearby mini PC (e.g., Intel NUC, Raspberry Pi). 
By leveraging this tool, \framework can measure DC power with a resolution of a few milliwatts and an accuracy in the order of 1.5\% for typical operating ranges.

Specifically, we developed a C++ 
measuring software\footnote{https://github.com/wineslab/YoctoWatt-RU-Power} that interfaces with the Yoctopuce high-level APIs to query the Yocto-Watt at high frequencies and obtain timestamped measurements of the \gls{ru} power draw throughout each experiment. Each API reading takes less than $1$\:$\mu$s, and samples are collected approximately every $60$\:ms (${\sim}16$\:sample/s). The timestamps allow us to align the \gls{ru} power traces with logs and metrics obtained from other tools and components on the OpenShift cluster.

\begin{figure}[h]
  \centering
  \includegraphics[width=\columnwidth]{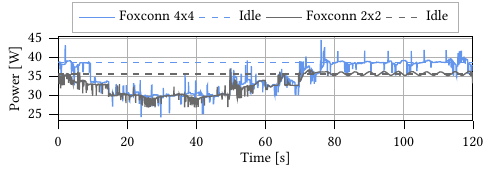}
  \caption{Foxconn RU power as measured by the Yocto-Watt.}
  \label{fig:yoctowatt-ru}
        \vspace{-.3cm}
\end{figure}

To provide an example of this, Figure~\ref{fig:yoctowatt-ru} shows the power trace measured of a Foxconn \gls{ru} as measured by the connected Yocto-Watt when the \gls{ru} is rebooted and configured to operate with 2$\times$2 and 4$\times$4 antenna layers.
We notice that the idle power stabilizes around $35.6$\:W in the 2$\times$2 configuration, and around $38.7$\:W in the 4$\times$4 one, showing that activating additional antenna elements increases the baseline power consumption of the \gls{ru}.

\subsection{5G Network Catalog} 
\label{sec:5g-testbed-components}

The remaining element in our \framework architecture is the 5G network catalog. This includes the software components that need to be deployed and tested to evaluate their energy consumption. 

In our prototype, the catalog includes multiple 5G core networks, \gls{ran} stacks, radios, and \glspl{ue} to support a wide range of experimental scenarios. On the core network side, we deploy two 5G cores---the open-source Open5GS as well as a commercial one---each in its own OpenShift namespace to keep workloads and configurations logically separated.
Their network functions (e.g., AMF, SMF, \gls{upf}) run as atomic pods, following a microservice-based approach.

We use the \gls{oai} and srsRAN software stacks for the \gls{ran}. Both stacks are open-source and run on general-purpose x86 servers in the OpenShift cluster used by \framework.
The \gls{cu} and \gls{du} functions are implemented in the same pod so that the pod-level power measurements reflect their combined consumption.
On the radio side, we leverage both commercial \glspl{ru} compliant with the O-RAN~7.2 functional split, and \gls{usrp} \glspl{sdr} operating with a split~8 configuration. This combination of commercial \glspl{ru} and \glspl{sdr} lets us evaluate energy and performance across heterogeneous hardware platforms while reusing the same core and \gls{ran} software components.

Finally, the \glspl{ue} in our testbed include Sierra Wireless EM9191 5G modems and commercial smartphones to support both scheduled experiments and mobility tests. Each modem is attached to a small host compute node (e.g., Raspberry Pi~5, Intel NUC, Lenovo ThinkCentre) positioned at designated locations.
At each location, a dedicated local switch connects these hosts and the nearby \glspl{ru} to the main switch of our OpenShift cluster through a fiber link. This physical layout allows us to test diverse \gls{ota} scenarios while keeping all core, \gls{ran}, and power-measurement components integrated into a single, automated framework.

\section{\framework Workflows}
\label{sec:architecture}

In this section, we describe the workflows implemented in \framework to profile the energy consumption of an \gls{e2e} O-RAN deployment in situ. The workflow provides a measurement scaffold that coordinates deployment configuration and synchronized collection of power and performance metrics across disaggregated O-RAN components. 

The architecture of \framework and its energy measurements pipeline are shown in Figure~\ref{fig:pipeline}. To profile the energy consumption of an \gls{e2e} O-RAN deployment, \framework executes the following four steps: (i) takes as input a high-level specification describing the network configuration to be tested (e.g., \gls{cu}/\gls{du}/\gls{ru} pair, core network, number of \glspl{ue}) as well as desired traffic conditions (e.g., requested bitrate in uplink and downlink for each user) and \gls{ric} applications to include in the test (e.g., \gls{kpm} xApps); (ii) configures and deploys network components for the test to be executed; (iii) collects synchronized power and network \glspl{kpm} from heterogeneous measurement sources for each component involved in the test using the hardware- and software-based measurement tools described in Section~\ref{sec:background}; and (iv) aggregates and stores measurement data for subsequent analysis. The following subsections describe each of these stages in more technical detail.
\subsection{Test Specifications}
\label{sec:test-specifications}

In \framework, tests are defined via \gls{json}-formatted test vectors, which specify the \textit{network scenario} and the \textit{traffic scenario} of the test. The former specifies the core network to use (e.g., Open5GS), the gNB protocol stack (e.g., \gls{oai}, srsRAN) together with its configuration parameters (e.g., numerology, frequency band, split, MIMO options), and \gls{ru} to use. The latter specifies the \glspl{ue} involved in the test and their traffic profile. This includes the test type (e.g., iPerf), protocol, target bitrate, direction, and duration, as well as the server endpoint for the test.

An example of testing vector is shown in Listing~\ref{lst:experiment-spec}. In this case, the test uses the commercial core network, an \gls{oai}-based \gls{cu}/\gls{du}, and a USRP \gls{sdr} as the radio device. The test vector also specifies a \gls{ue} running an iPerf \gls{udp} test at the target rate of $70$\:Mbps for $60$\:s in the downlink direction (``reverse'' option in the test vector). This machine-readable specification is then parsed by \framework automation pipelines to deploy the requested components and configure the traffic generator accordingly.
\begin{lstlisting}[float=ht,floatplacement=h,language=json,style=mystyle-json, 
caption={Test vector example.}, 
label={lst:experiment-spec}]
{
  "network_scenario": {
    "id": 1,
    "core_network": {
      "name": "commercial"
    },
    "ran": {
      "cu": {
        "name": "oai-cu",
        "config_file": "oai_cu.conf"
      },
      "du": {
        "name": "oai-du",
        "config_file": "oai_162prb.conf"
      },
      "functional_split": "8",
      "ru": {
        "name": "usrp",
        "address": "192.168.40.20"
      }
    }
  },
  "traffic_scenario": {
    "id": 1,
    "ue_specification": [
      {
        "slice_id": 1,
        "test_type": "iperf",
        "bandwidth_mbps": 70,
        "duration": 60,
        "protocol": "udp",
        "reverse": true,
        "json_output": true,
        "server_hostname": "<url-hostname>",
        "server_port": 32205
      }
    ]
  }
}

\end{lstlisting}

\subsection{Deployment and Execution of the Test}
\label{sec:deployment-execution}

After tests have been specified via dedicated test vectors, their deployment and execution are orchestrated by Tekton pipelines running on our OpenShift cluster.
These read the \gls{json} experiment specification (see Listing~\ref{lst:experiment-spec}), extracts the selected core network, \gls{cu}/\gls{du} stack, functional split option, \gls{ru}, and traffic parameters, and executes a sequence of tasks to deploy and connect the required components. These tasks include deploying the gNB protocol stack pods with the specified configuration, configuring interfaces toward the selected \gls{ru} and core network, and starting the gNB process.
Pipelines are synchronized and version-controlled via ArgoCD to ensure that they can be seamlessly ported and executed on other testbeds.

During the test execution, information such as the start time of the test is saved and used to align performance trends with the energy results returned by the energy measurement tools of Section~\ref{sec:energy-measurement-tools} (e.g., Kepler, metered \glspl{pdu}, Yocto-Watt). Similarly, the traffic scenario is sent to the selected \glspl{ue} (Sierra Wireless 5G modems in our case), which attach to the running gNB and start the iPerf-based test with the requested parameters.
When the test completes, \framework records the end timestamp together with the performance metrics reported by the \glspl{ue}. These metadata are then used by the power-collector and data-collector components (Section~\ref{sec:data-collection}) to analyze the power and performance measurements.

\subsection{Data Collection}
\label{sec:data-collection}

After each test, \framework uses two dedicated pods to collect measurement data: a \emph{power collector} and a \emph{performance collector} pod (see Figure~\ref{fig:pipeline}). These expose HTTP Flask API and are leveraged to receive the test results to analyze. Specifically, the power collector pod receives metadata describing the executed test, start and end timestamps, selected network and traffic scenarios, and the name of the file where the Yocto-Watt traces were saved on the companion PC.
Using this metadata, it queries Kepler for the pod-level energy metrics in the specified timestamps (e.g., for the core network and gNB pods). Then, it fetches the compute node-level Raritan PX4 measurements for the node where the pods used in the test ran, as well as the \gls{ru} power trace from the mini PC connected to the Yocto-Watt module. All power measurements are then combined and stored in a central database.

The performance collector pod, instead, aggregates the results of the traffic tests and gNB logs. In particular, it receives the iPerf test results from the \glspl{ue} involved in the test. Data is stored in a central database alongside the relevant experiment metadata so that it can be aligned with the power traces gathered by the power collector pod.
The dataset formed in this way lets us derive energy-efficiency metrics for specific deployments, and to compare the performance and energy consumption of different configurations and traffic scenarios.

\section{Experimental Results}
\label{sec:experiments}
In this section, we present results obtained via \gls{e2e} \gls{ota} experiments conducted using \framework. Section~\ref{sec:oai-srsran-comparison} compares the power consumption of
5G \gls{ran} protocol stacks under different traffic loads. Section~\ref{sec:core-network-power} focuses on the power consumption of the core network. Finally, Section~\ref{sec:e2e-power} discusses the power consumption of individual components in \gls{e2e} experiments with xApps in the loop.
%

\subsection{RAN Measurements}
\label{sec:oai-srsran-comparison}
This section presents experimental results on the power consumption of two different \gls{ran} protocol stacks, \gls{oai} (version 2025.w36) and srsRAN (version 25.04). Both stacks use a split-8 using an Ettus Research X410 \glspl{usrp} as the \gls{ru}. All \gls{e2e} experiments in this section are performed by deploying the gNB pod on a Microway server and a Sierra 5G modem as the \gls{ue}. The power consumption of the pods is measured with Kepler through a series of automated \gls{ota} experiments.

\begin{figure}[t]
  \centering
  \begin{subfigure}[a]{\columnwidth}
    \centering
    \includegraphics[width=\columnwidth]{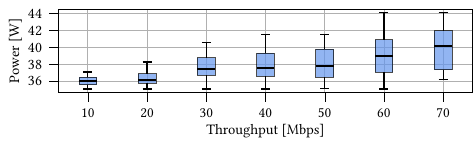}
    \caption{\gls{oai} protocol stack}
    \label{fig:udp-oai}
  \end{subfigure}
  \vspace{0.1cm}
  \begin{subfigure}[b]{\columnwidth}
    \centering
    \includegraphics[width=\columnwidth]{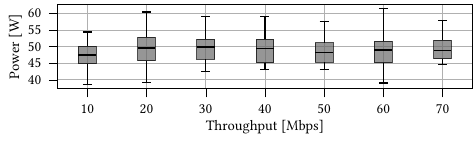}
    \caption{srsRAN protocol stack}
    \label{fig:udp-srsran}
  \end{subfigure}
  \caption{gNB power measurements under different \gls{udp} traffic loads.}
  \label{fig:udp-comparison}
  \vspace{-.3cm}
\end{figure}

Figure~\ref{fig:udp-comparison} reports experiments with different \gls{udp} traffic loads from $10$ to $70$\:Mbps between the gNB and the \gls{ue}. Figure~\ref{fig:udp-oai} shows box plots of the power consumption of the \gls{oai} gNB pod, while Figure~\ref{fig:udp-srsran} shows the corresponding box plots for the srsRAN gNB pod under the same traffic loads.
The \gls{oai} pod exhibits a gradual, approximately linear increase in consumption from about $36$ to $40$\:W as the load increases, whereas the srsRAN pod consistently consumes around $48$\:W for all loads.
These results suggest that certain software stacks (e.g., \gls{oai}) show load-dependent energy consumption, while that of others (e.g., srsRAN) does not depend as heavily from the load. While the nature of such differences is hard to characterize---and beyond the scope of this paper---we notice that \gls{oai} and srsRAN are implemented in two different programming languages. The former is predominantly C-based, which is a low-level programming language, while the latter is implemented C++, which leverages higher-layer programming directives that may introduce additional overhead~\cite{kempen_its_2025}.

\begin{figure}[t]
  \centering
  \begin{subfigure}[b]{0.48\columnwidth}
    \centering
    \includegraphics[width=\textwidth]{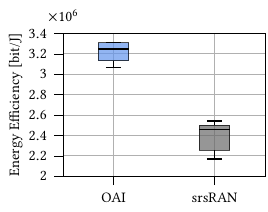}
    \caption{Downlink}
    \label{fig:downlink-tcp-energy-efficiency}
  \end{subfigure}
  \hfill
  \begin{subfigure}[b]{0.48\columnwidth}
    \centering
    \includegraphics[width=\textwidth]{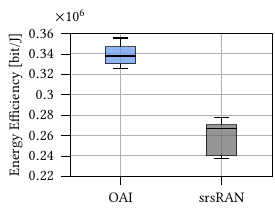}
    \caption{Uplink}
    \label{fig:uplink-tcp-energy-efficiency}
  \end{subfigure}
  \caption{Energy efficiency for \gls{oai} and srsRAN.}
  \label{fig:tcp-energy-efficiency}
  \vspace{-.3cm}
\end{figure}

To better understand the energy efficiency of different stacks and configurations, we also compare the energy efficiency for the \gls{oai} and srsRAN protocol stacks under \gls{tcp} traffic loads. Figure~\ref{fig:tcp-energy-efficiency} shows this metric for both \gls{oai} and srsRAN in the downlink (Figure~\ref{fig:downlink-tcp-energy-efficiency}) and uplink (Figure~\ref{fig:uplink-tcp-energy-efficiency}) directions.
In Figure~\ref{fig:downlink-tcp-energy-efficiency}, $95\%$ confidence intervals of the downlink energy efficieny of the \gls{oai} pod range from $3.02$ to $3.53$\:Mbit/J, while those of the srsRAN pod are between $2.18$ and $2.59$\:Mbit/J.
In the uplink direction, shown in Figure~\ref{fig:uplink-tcp-energy-efficiency}, confidence intervals are between $0.318$ and $0.360$\:Mbit/J for \gls{oai}, and between $0.236$ and $0.282$\:Mbit/J for srsRAN. These results confirm that srsRAN consumes more energy per bit than \gls{oai} in both directions. They also show that uplink transmissions are approximately an order of magnitude more energy demanding than downlink, highlighting the importance of considering traffic direction when evaluating the energy efficiency of the \gls{ran}.

%

\subsection{Core Network Measurements}
\label{sec:core-network-power}
\begin{figure}[h]
  \centering
  \includegraphics[width=\columnwidth]{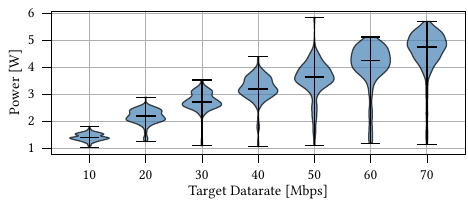}
  \caption{Power consumption of the core network UPF pod under different UDP loads.}
  \label{fig:core-power}
        \vspace{-.3cm}
\end{figure}

In this section, we focus on the power consumption of the core network.
Specifically, we focus on the \gls{upf} pod since it is the main data-plane component responsible for forwarding all user traffic between the \gls{ran} and the external data network. As a result, different loads impact power consumption differently. We leverage Kepler to obtain these measurements. Figure~\ref{fig:core-power} shows a violin plot of the power consumption of the commercial core network \gls{upf} while exchanging \gls{udp} traffic for various deployments of gNBs and \glspl{ue}. Other core interfaces are not shown, as their power consumption is negligible compared to the \gls{upf} and does not vary with the traffic load. As can be observed, the \gls{upf} power consumption increases almost linearly from about $1.5$ to $5$\:W as the traffic load increases from $10$ to $70$\:Mbps.

\subsection{End-to-End Measurements}
\label{sec:e2e-power}
In this section, we present an \gls{ota} \gls{e2e} experiment designed to collect power measurements from the \gls{cu}/\gls{du}, \gls{ru}, and xApp while increasing the number of attached \glspl{ue}. For these experiments, we rely on multiple smartphones (Samsung Galaxy S23, OnePlus AC Nord 2003, Google Pixel).
We evaluate two different monitoring xApps: (i) the first one is based on a lightweight, non standard-compliant protobuf-based E2SM implementation;\footnote{https://github.com/wineslab/xapp-oai/tree/kpm-xapp} and (ii) the second is fully O-RAN compliant and follows standard E2AP and E2SM O-RAN specifications.\footnote{https://github.com/wineslab/xDevSM-xapps-examples} Both xApps implement a basic \gls{kpm} monitoring tool sending data to an external database.

The \gls{cu}/\gls{du} pod is instantiated on a Dell R760 server and runs the \gls{oai} 5G gNB stack (version 2024.w51) and connects to a commercial Foxconn \gls{ru} using the 7.2 functional split.
The power of the virtualized pods, including the \gls{cu}/\gls{du} and the xApp, is measured with Kepler, while the power consumption of the \gls{ru} is measured using the Yocto-Watt board.

Table~\ref{tab:e2e-power} shows the average power consumption of the \gls{cu}/\gls{du} pod and of the \gls{ru} for experiments with different numbers of attached \glspl{ue} when running the two xApps. The \gls{cu}/\gls{du} consumption increases approximately linearly with the number of users, from approximately $11$ to $26$\:W, reflecting the higher processing load as more \glspl{ue} are served. In contrast, the \gls{ru} power remains nearly constant at around $35$\:W, indicating that in this setup the radio hardware consumption is only loosely dependent on the number of active users.

\begin{table}[tb]
    \centering
    \footnotesize
    \setlength{\tabcolsep}{3pt}
    \caption{Average of gNB power consumption under different loads when using the protobuf and standard-compliant xApps.}
    \label{tab:e2e-power}
\begin{tabularx}{\columnwidth}{
    >{\hsize=1.2\hsize\centering\arraybackslash}X  
    >{\hsize=1.05\hsize\centering\arraybackslash}X 
    >{\hsize=0.85\hsize\centering\arraybackslash}X 
    >{\hsize=1.05\hsize\centering\arraybackslash}X 
    >{\hsize=0.85\hsize\centering\arraybackslash}X 
}
        \toprule
        \multirow{2}{*}{\parbox{1.5cm}{\centering Number of \\ users}} & \multicolumn{2}{c}{ Protobuf xApp} & \multicolumn{2}{c}{Standard-compliant xApp} \\
        \cmidrule(lr){2-3} \cmidrule(lr){4-5}
        & CU/DU [W] & RU [W] & CU/DU [W] & RU [W] \\
        \midrule
        1 & 11.9 & 35.3606 & 10.95 & 35.2991 \\
        2 & 13.2 & 35.4810 & 11.65 & 35.4462 \\
        3 & 24.35 & 35.5110 & 22.55 & 35.5937 \\
        4 & 26.15 & 35.5836 & 25.9 & 35.6625 \\
        5 & 26.3 & 35.6025 & 26.15 & 35.6925 \\
        \bottomrule
    \end{tabularx}
\end{table}

\begin{figure}[h]
  \centering
  \includegraphics[width=\columnwidth]{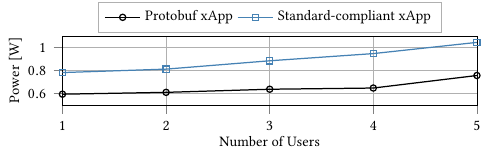}
  \caption{xApp pod power measurements under different loads.}
  \label{fig:xapp-power}
        \vspace{-.3cm}
\end{figure}

Figure~\ref{fig:xapp-power} shows the power consumption of the two xApps as the number of \glspl{ue} increases. Both xApps implement the E2SM-KPM service model and periodically extract and report \glspl{kpm} for data visualization and performance monitoring. Despite this relatively lightweight functionality, the power measured by Kepler increases with the number of users, and, due to their different implementations, the standard-compliant xApp consistently consumes more power than the protobuf-based xApp under the same load.

\begin{figure}[h]
  \centering
  \includegraphics[width=\columnwidth]{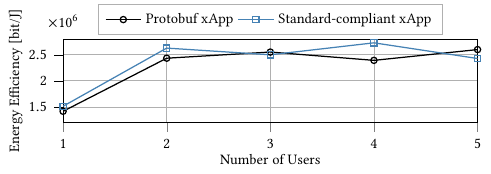}
  \caption{Energy efficiency of the network for two xApps.}
  \label{fig:energy-efficiency-xapp}
\end{figure}

Figure~\ref{fig:energy-efficiency-xapp} presents energy efficiency,
computed as the aggregate number of bits received by the \glspl{ue} divided by the total energy consumption (including \gls{cu}, \gls{du}, \gls{ru}, and the xApp). We notice that the energy efficiency increases as the number of users increases. This highlights the effect of infrastructure sharing from an energy perspective: when a single user is active, the total bits are on average about $1.5$\:Mbit per consumed~Joule, whereas with more users this value increases to approximately $2.5$\:Mbit per consumed~Joule.

\section{Conclusions}
\label{sec:conclusion}
This paper presented \framework, an automated framework for end-to-end energy and performance profiling of O-RAN deployment. We prototyped \framework on a Red Hat OpenShift cluster integrating multiple core networks, \gls{ran} protocol stacks, radios, and \glspl{ue} together with Raritan PX4, Kepler, and Yocto-Watt measurement tools. Our experiments show that different \gls{ran} stacks exhibit distinct energy-efficiency behaviors, as well as different consumption in uplink and downlink directions. We also showed how the core network power consumption scales with load, and end-to-end energy profiling with different xApps in the loop. Overall, \framework provides a practical foundation for systematic, data-driven exploration of energy–performance trade-offs in O-RAN deployments and for future work on energy-aware \gls{ric} applications.

\balance
\bibliographystyle{IEEEtran}
\bibliography{biblio}

\end{document}

%% file: acronyms.tex
\newacronym{3gpp}{3GPP}{3rd Generation Partnership Project}
\newacronym{4g}{4G}{4th generation mobile network}
\newacronym{5g}{5G}{5th generation mobile network}
\newacronym{6g}{6G}{6th generation mobile network}
\newacronym{nextg}{NextG}{Next Generation}
\newacronym{5gc}{5GC}{5G Core}
\newacronym{acpi}{ACPI}{Advanced Configuration and Power Interface}
\newacronym{adc}{ADC}{Analog to Digital Converter}
\newacronym{aerpaw}{AERPAW}{Aerial Experimentation and Research Platform for Advanced Wireless}
\newacronym{ai}{AI}{Artificial Intelligence}
\newacronym{aimd}{AIMD}{Additive Increase Multiplicative Decrease}
\newacronym{ict}{ICT}{Information and Communication Technology}
\newacronym{am}{AM}{Acknowledged Mode}
\newacronym{amc}{AMC}{Adaptive Modulation and Coding}
\newacronym{amf}{AMF}{Access and Mobility Management Function}
\newacronym{aops}{AOPS}{Adaptive Order Prediction Scheduling}
\newacronym{api}{API}{Application Programming Interface}
\newacronym{apn}{APN}{Access Point Name}
\newacronym{aqm}{AQM}{Active Queue Management}
\newacronym{ausf}{AUSF}{Authentication Server Function}
\newacronym{avc}{AVC}{Advanced Video Coding}
\newacronym{awgn}{AGWN}{Additive White Gaussian Noise}
\newacronym{balia}{BALIA}{Balanced Link Adaptation Algorithm}
\newacronym{bbu}{BBU}{Base Band Unit}
\newacronym{bdp}{BDP}{Bandwidth-Delay Product}
\newacronym{ber}{BER}{Bit Error Rate}
\newacronym{bf}{BF}{Beamforming}
\newacronym{bler}{BLER}{Block Error Rate}
\newacronym{brr}{BRR}{Bayesian Ridge Regressor}
\newacronym{bsr}{BSR}{Buffer Status Report}
\newacronym{bs}{BS}{Base Station}
\newacronym{bpsk}{BPSK}{Binary Phase-shift keying}
\newacronym{bss}{BSS}{Business Support System}
\newacronym{ca}{CA}{Carrier Aggregation}
\newacronym{caas}{CaaS}{Connectivity-as-a-Service}
\newacronym{cb}{CB}{Code Block}
\newacronym{cc}{CC}{Congestion Control}
\newacronym{ccid}{CCID}{Congestion Control ID}
\newacronym{cco}{CC}{Carrier Component}
\newacronym{cd}{CD}{Continuous Delivery}
\newacronym{cdd}{CDD}{Cyclic Delay Diversity}
\newacronym{cdf}{CDF}{Cumulative Distribution Function}
\newacronym{cdma}{CDMA}{Code-Division Multiple Access}
\newacronym{cdn}{CDN}{Content Distribution Network}
\newacronym{ci}{CI}{Continuous Integration}
\newacronym{cicd}{CI/CD}{Continuous Integration/Continuous Delivery}
\newacronym{cir}{CIR}{Channel Impulse Response}
\newacronym{cn}{CN}{Core Network}
\newacronym{cncf}{CNCF}{Cloud Native Computing Foundation}
\newacronym{codel}{CoDel}{Controlled Delay Management}
\newacronym{comac}{COMAC}{Converged Multi-Access and Core}
\newacronym{cord}{CORD}{Central Office Re-architected as a Datacenter}
\newacronym{cornet}{CORNET}{COgnitive Radio NETwork}
\newacronym{cosmos}{COSMOS}{Cloud Enhanced Open Software Defined Mobile Wireless Testbed for City-Scale Deployment}
\newacronym{cots}{COTS}{Commercial Off-the-Shelf}
\newacronym{cp}{CP}{Control Plane}
\newacronym{cpu}{CPU}{Central Processing Unit}
\newacronym{cqi}{CQI}{Channel Quality Information}
\newacronym{cr}{CR}{Cognitive Radio}
\newacronym{cran}{CRAN}{Cloud \gls{ran}}
\newacronym{crs}{CRS}{Cell Reference Signal}
\newacronym{csi}{CSI}{Channel State Information}
\newacronym{csirs}{CSI-RS}{Channel State Information - Reference Signal}
\newacronym{cu}{CU}{Central Unit}
\newacronym{cups}{CUPS}{Control Plane and User Plane Separation}
\newacronym{d2tcp}{D$^2$TCP}{Deadline-aware Data center TCP}
\newacronym{d3}{D$^3$}{Deadline-Driven Delivery}
\newacronym{dac}{DAC}{Digital to Analog Converter}
\newacronym{dag}{DAG}{Directed Acyclic Graph}
\newacronym{darpa}{DARPA}{Defense Advanced Research Projects Agency}
\newacronym{das}{DAS}{Distributed Antenna System}
\newacronym{dash}{DASH}{Dynamic Adaptive Streaming over HTTP}
\newacronym{dc}{DC}{Dual Connectivity}
\newacronym{dccp}{DCCP}{Datagram Congestion Control Protocol}
\newacronym{dce}{DCE}{Direct Code Execution}
\newacronym{dci}{DCI}{Downlink Control Information}
\newacronym{dcl}{DCL}{Dear Colleague Letter}
\newacronym{dctcp}{DCTCP}{Data Center TCP}
\newacronym{devops}{DevOps}{Development and Operations}
\newacronym{dl}{DL}{Deep Learning}
\newacronym{dmr}{DMR}{Deadline Miss Ratio}
\newacronym{dmrs}{DMRS}{DeModulation Reference Signal}
\newacronym{drlcc}{DRL-CC}{Deep Reinforcement Learning Congestion Control}
\newacronym{drs}{DRS}{Discovery Reference Signal}
\newacronym{dt}{DT}{Digital Twin}
\newacronym{dtn}{DTN}{Digital Twin Network}
\newacronym{dtmn}{DTMN}{Digital Twin for Mobile Network}
\newacronym{dtwn}{DTWN}{Digital Twin Wireless Network}
\newacronym{du}{DU}{Distributed Unit}
\newacronym{e2e}{E2E}{end-to-end}
\newacronym{ebpf}{eBPF}{extended Berkeley Packet Filter}
\newacronym{ecaas}{ECaaS}{Edge-Cloud-as-a-Service}
\newacronym{ecn}{ECN}{Explicit Congestion Notification}
\newacronym{edf}{EDF}{Earliest Deadline First}
\newacronym{em}{EM}{Electro-Magnetic}
\newacronym{embb}{eMBB}{Enhanced Mobile Broadband}
\newacronym{empower}{EMPOWER}{EMpowering transatlantic PlatfOrms for advanced WirEless Research}
\newacronym{enb}{eNB}{evolved Node Base}
\newacronym{endc}{EN-DC}{E-UTRAN-\gls{nr} \gls{dc}}
\newacronym{epc}{EPC}{Evolved Packet Core}
\newacronym{eps}{EPS}{Evolved Packet System}
\newacronym{es}{ES}{Edge Server}
\newacronym{etsi}{ETSI}{European Telecommunications Standards Institute}
\newacronym[firstplural=Estimated Times of Arrival (ETAs)]{eta}{ETA}{Estimated Time of Arrival}
\newacronym{eutran}{E-UTRAN}{Evolved Universal Terrestrial Access Network}
\newacronym{faas}{FaaS}{Function-as-a-Service}
\newacronym{fapi}{FAPI}{Functional Application Platform Interface}
\newacronym{fcc}{FCC}{Federal Communications Commission}
\newacronym{fdd}{FDD}{Frequency Division Duplexing}
\newacronym{fdm}{FDM}{Frequency Division Multiplexing}
\newacronym{fdma}{FDMA}{Frequency Division Multiple Access}
\newacronym{fed4fire}{FED4FIRE+}{Federation 4 Future Internet Research and Experimentation Plus}
\newacronym{fir}{FIR}{Finite Impulse Response}
\newacronym{fit}{FIT}{Future \acrlong{iot}}
\newacronym{fl}{FL}{Federated Learning}
\newacronym{fpga}{FPGA}{Field Programmable Gate Array}
\newacronym{fr2}{FR2}{Frequency Range 2}
\newacronym{fs}{FS}{Fast Switching}
\newacronym{fscc}{FSCC}{Flow Sharing Congestion Control}
\newacronym{ftp}{FTP}{File Transfer Protocol}
\newacronym{fw}{FW}{Flow Window}
\newacronym{ga128}{Ga}{Golay Sequence type A}
\newacronym{ge}{GE}{Gaussian Elimination}
\newacronym{glfsr}{GLFSR}{Galois Linear Feedback Shift Register}
\newacronym{gnb}{gNB}{Next Generation Node Base}
\newacronym{gold}{Gold}{Gold}
\newacronym{gop}{GOP}{Group of Pictures}
\newacronym{gpr}{GPR}{Gaussian Process Regressor}
\newacronym{gpu}{GPU}{Graphics Processing Unit}
\newacronym{gtp}{GTP}{GPRS Tunneling Protocol}
\newacronym{gtpc}{GTP-C}{GPRS Tunnelling Protocol Control Plane}
\newacronym{gtpu}{GTP-U}{GPRS Tunnelling Protocol User Plane}
\newacronym{gtpv2c}{GTPv2-C}{\gls{gtp} v2 - Control}
\newacronym{gw}{GW}{Gateway}
\newacronym{harq}{HARQ}{Hybrid Automatic Repeat reQuest}
\newacronym{hetnet}{HetNet}{Heterogeneous Network}
\newacronym{hh}{HH}{Hard Handover}
\newacronym{hol}{HOL}{Head-of-Line}
\newacronym{hqf}{HQF}{Highest-quality-first}
\newacronym{hss}{HSS}{Home Subscription Server}
\newacronym{http}{HTTP}{HyperText Transfer Protocol}
\newacronym{ia}{IA}{Initial Access}
\newacronym{iab}{IAB}{Integrated Access and Backhaul}
\newacronym{ic}{IC}{Incident Command}
\newacronym{idrac}{iDRAC}{Integrated Dell Remote Access Controller}
\newacronym{ietf}{IETF}{Internet Engineering Task Force}
\newacronym{ifw}{IFW}{Interference Free Window}
\newacronym{imsi}{IMSI}{International Mobile Subscriber Identity}
\newacronym{imt}{IMT}{International Mobile Telecommunication}
\newacronym{iot}{IoT}{Internet of Things}
\newacronym{ip}{IP}{Internet Protocol}
\newacronym{iq}{IQ}{In-phase and Quadrature}
\newacronym{isi}{ISI}{Inter-Symbol Interference}
\newacronym{itu}{ITU}{International Telecommunication Union}
\newacronym{json}{JSON}{JavaScript Object Notation}
\newacronym{kpi}{KPI}{Key Performance Indicator}
\newacronym{kpm}{KPM}{Key Performance Measurement}
\newacronym{kvm}{KVM}{Kernel-based Virtual Machine}
\newacronym{lfsr}{LFSR}{Linear Feedback Shift Register}
\newacronym{los}{LOS}{Line-of-Sight}
\newacronym{ls}{LS}{Loosely Synchronised}
\newacronym{lsm}{LSM}{Link-to-System Mapping}
\newacronym{lstm}{LSTM}{Long Short Term Memory}
\newacronym{lte}{LTE}{Long Term Evolution}
\newacronym{lxc}{LXC}{Linux Container}
\newacronym{m2m}{M2M}{Machine to Machine}
\newacronym{mac}{MAC}{Medium Access Control}
\newacronym{mai}{MAI}{Multiple Access Interference}
\newacronym{manet}{MANET}{Mobile Ad Hoc Network}
\newacronym{mano}{MANO}{Management and Orchestration}
\newacronym{mc}{MC}{Multi-Connectivity}
\newacronym{mcc}{MCC}{Mobile Cloud Computing}
\newacronym{mchem}{MCHEM}{Massive Channel Emulator}
\newacronym{mcs}{MCS}{Modulation and Coding Scheme}
\newacronym{mec}{MEC}{Multi-access Edge Computing}
\newacronym{mec2}{MEC}{Mobile Edge Cloud}
\newacronym{mec3}{MEC}{Mobile Edge Computing}
\newacronym{mfc}{MFC}{Mobile Fog Computing}
\newacronym{mi}{MI}{Mutual Information}
\newacronym{mib}{MIB}{Master Information Block}
\newacronym{miesm}{MIESM}{Mutual Information Based Effective SINR}
\newacronym{mimo}{MIMO}{Multiple Input, Multiple Output}
\newacronym{mgen}{MGEN}{Multi-Generator}
\newacronym{ml}{ML}{Machine Learning}
\newacronym{mlr}{MLR}{Maximum-local-rate}
\newacronym[plural=\gls{mme}s,firstplural=Mobility Management Entities (MMEs)]{mme}{MME}{Mobility Management Entity}
\newacronym{mmtc}{mMTC}{Massive Machine-Type Communications}
\newacronym{mmwave}{mmWave}{millimeter wave}
\newacronym{mpdccp}{MP-DCCP}{Multipath Datagram Congestion Control Protocol}
\newacronym{mptcp}{MPTCP}{Multipath TCP}
\newacronym{mr}{MR}{Maximum Rate}
\newacronym{mrdc}{MR-DC}{Multi \gls{rat} \gls{dc}}
\newacronym{mse}{MSE}{Mean Square Error}
\newacronym{mss}{MSS}{Maximum Segment Size}
\newacronym{mt}{MT}{Mobile Termination}
\newacronym{mtd}{MTD}{Machine-Type Device}
\newacronym{mtu}{MTU}{Maximum Transmission Unit}
\newacronym{mumimo}{MU-MIMO}{Multi-user \gls{mimo}}
\newacronym{mvno}{MVNO}{Mobile Virtual Network Operator}
\newacronym{nalu}{NALU}{Network Abstraction Layer Unit}
\newacronym{nas}{NAS}{Network Attached Storage}
\newacronym{nbiot}{NB-IoT}{Narrow Band IoT}
\newacronym{nfv}{NFV}{Network Function Virtualization}
\newacronym{nfvi}{NFVI}{Network Function Virtualization Infrastructure}
\newacronym{nic}{NIC}{Network Interface Card}
\newacronym{nlos}{NLOS}{Non-Line-of-Sight}
\newacronym{now}{NOW}{Non Overlapping Window}
\newacronym{nrdz}{NRDZ}{National Radio Dynamic Zone}
\newacronym{nsf}{NSF}{National Science Foundation}
\newacronym{nsm}{NSM}{Network Service Mesh}
\newacronym[type=hidden]{nr}{NR}{New Radio}
\newacronym{nrf}{NRF}{Network Repository Function}
\newacronym{nsa}{NSA}{Non Stand Alone}
\newacronym{nse}{NSE}{Network Slicing Engine}
\newacronym{nssf}{NSSF}{Network Slice Selection Function}
\newacronym{ntp}{NTP}{Network Time Protocol}
\newacronym{nvml}{NVML}{NVIDIA Management Library}
\newacronym{o2i}{O2I}{Outdoor to Indoor}
\newacronym{oai}{OAI}{OpenAirInterface}
\newacronym{oaicn}{OAI-CN}{\gls{oai} \acrlong{cn}}
\newacronym{oairan}{OAI-RAN}{\acrlong{oai} \acrlong{ran}}
\newacronym{oam}{OAM}{Operations, Administration and Maintenance}
\newacronym[plural=\gls{obu}s,firstplural=Onboard Units (OBUs)]{obu}{OBU}{Onboard Unit}
\newacronym{ofdm}{OFDM}{Orthogonal Frequency Division Multiplexing}
\newacronym{olia}{OLIA}{Opportunistic Linked Increase Algorithm}
\newacronym{omec}{OMEC}{Open Mobile Evolved Core}
\newacronym{onap}{ONAP}{Open Network Automation Platform}
\newacronym{onf}{ONF}{Open Networking Foundation}
\newacronym{onos}{ONOS}{Open Networking Operating System}
\newacronym{oom}{OOM}{\gls{onap} Operations Manager}
\newacronym{opnfv}{OPNFV}{Open Platform for \gls{nfv}}
\newacronym{opex}{OPEX}{Operating Expenditure}
\newacronym[type=hidden]{oran}{O-RAN}{Open \gls{ran}}
\newacronym{orbit}{ORBIT}{Open-Access Research Testbed for Next-Generation Wireless Networks}
\newacronym{os}{OS}{Operating System}
\newacronym{osm}{OSM}{Open Street Map}
\newacronym{oss}{OSS}{Operations Support System}
\newacronym{pa}{PA}{Position-aware}
\newacronym{pase}{PASE}{Prioritization, Arbitration, and Self-adjusting Endpoints}
\newacronym{pawr}{PAWR}{Platforms for Advanced Wireless Research}
\newacronym{pbch}{PBCH}{Physical Broadcast Channel}
\newacronym{pcef}{PCEF}{Policy and Charging Enforcement Function}
\newacronym{pcfich}{PCFICH}{Physical Control Format Indicator Channel}
\newacronym{pcrf}{PCRF}{Policy and Charging Rules Function}
\newacronym{pdcch}{PDCCH}{Physical Downlink Control Channel}
\newacronym{pdcp}{PDCP}{Packet Data Convergence Protocol}
\newacronym{pdsch}{PDSCH}{Physical Downlink Shared Channel}
\newacronym{pdu}{PDU}{Power Distribution Unit}
\newacronym{pdp}{PDP}{Power Delay Profile}
\newacronym{pf}{PF}{Proportional Fair}
\newacronym{pgw}{PGW}{Packet Gateway}
\newacronym{phich}{PHICH}{Physical Hybrid ARQ Indicator Channel}
\newacronym{phy}{PHY}{Physical}
\newacronym{pl}{PL}{Path Loss}
\newacronym{pmch}{PMCH}{Physical Multicast Channel}
\newacronym{pmi}{PMI}{Precoding Matrix Indicators}
\newacronym{powder}{POWDER}{Platform for Open Wireless Data-driven Experimental Research}
\newacronym{ppo}{PPO}{Proximal Policy Optimization}
\newacronym{ppp}{PPP}{Poisson Point Process}
\newacronym{prach}{PRACH}{Physical Random Access Channel}
\newacronym{prb}{PRB}{Physical Resource Block}
\newacronym{psnr}{PSNR}{Peak Signal to Noise Ratio}
\newacronym{pss}{PSS}{Primary Synchronization Signal}
\newacronym{pucch}{PUCCH}{Physical Uplink Control Channel}
\newacronym{pusch}{PUSCH}{Physical Uplink Shared Channel}
\newacronym{qam}{QAM}{Quadrature Amplitude Modulation}
\newacronym{qci}{QCI}{\gls{qos} Class Identifier}
\newacronym{qoe}{QoE}{Quality of Experience}
\newacronym{qos}{QoS}{Quality of Service}
\newacronym{qtgui}{QT-GUI}{QT Graphical User Interface}
\newacronym{quic}{QUIC}{Quick UDP Internet Connections}
\newacronym{rach}{RACH}{Random Access Channel}
\newacronym{ran}{RAN}{Radio Access Network}
\newacronym[firstplural=Radio Access Technologies (RATs)]{rat}{RAT}{Radio Access Technology}
\newacronym{rapl}{RAPL}{Running Average Power Limit}
\newacronym{rcn}{RCN}{Research Coordination Network}
\newacronym{rec}{REC}{Radio Edge Cloud}
\newacronym{red}{RED}{Random Early Detection}
\newacronym{renew}{RENEW}{Reconfigurable Eco-system for Next-generation  Wireless}
\newacronym{rf}{RF}{Radio Frequency}
\newacronym{rfc}{RFC}{Request for Comments}
\newacronym{rfr}{RFR}{Random Forest Regressor}
\newacronym{ric}{RIC}{RAN Intelligent Controller}
\newacronym{rlc}{RLC}{Radio Link Control}
\newacronym{rlf}{RLF}{Radio Link Failure}
\newacronym{rlnc}{RLNC}{Random Linear Network Coding}
\newacronym{rmse}{RMSE}{Root Mean Squared Error}
\newacronym{rnis}{RNIS}{Radio Network Information Service}
\newacronym{rr}{RR}{Round Robin}
\newacronym{rrc}{RRC}{Radio Resource Control}
\newacronym{rrm}{RRM}{Radio Resource Management}
\newacronym{rru}{RRU}{Remote Radio Unit}
\newacronym{rs}{RS}{Remote Server}
\newacronym{rsrp}{RSRP}{Reference Signal Received Power}
\newacronym{rsrq}{RSRQ}{Reference Signal Received Quality}
\newacronym{rss}{RSS}{Received Signal Strength}
\newacronym{rssi}{RSSI}{Received Signal Strength Indicator}
\newacronym{rsu}{RSU}{Road-Side Unit}
\newacronym{rtt}{RTT}{Round Trip Time}
\newacronym{ru}{RU}{Radio Unit}
\newacronym{rw}{RW}{Receive Window}
\newacronym{rx}{RX}{Receiver}
\newacronym{s1ap}{S1AP}{S1 Application Protocol}
\newacronym{sa}{SA}{standalone}
\newacronym{sack}{SACK}{Selective Acknowledgment}
\newacronym{sap}{SAP}{Service Access Point}
\newacronym{sc2}{SC2}{Spectrum Collaboration Challenge}
\newacronym{scef}{SCEF}{Service Capability Exposure Function}
\newacronym{sch}{SCH}{Secondary Cell Handover}
\newacronym{scoot}{SCOOT}{Split Cycle Offset Optimization Technique}
\newacronym{sctp}{SCTP}{Stream Control Transmission Protocol}
\newacronym{sdap}{SDAP}{Service Data Adaptation Protocol}
\newacronym{sd}{SD}{Standard Deviation}
\newacronym{sdk}{SDK}{Software Development Kit}
\newacronym{sdm}{SDM}{Space Division Multiplexing}
\newacronym{sdma}{SDMA}{Spatial Division Multiple Access}
\newacronym{sdn}{SDN}{Software-defined Networking}
\newacronym{sdr}{SDR}{Software-defined Radio}
\newacronym{seba}{SEBA}{SDN-Enabled Broadband Access}
\newacronym{sgsn}{SGSN}{Serving GPRS Support Node}
\newacronym{sgw}{SGW}{Service Gateway}
\newacronym{si}{SI}{Study Item}
\newacronym{sib}{SIB}{Secondary Information Block}
\newacronym{sinr}{SINR}{Signal to Interference plus Noise Ratio}
\newacronym{sip}{SIP}{Session Initiation Protocol}
\newacronym{siso}{SISO}{Single Input, Single Output}
\newacronym{sla}{SLA}{Service Level Agreement}
\newacronym{sm}{SM}{Saturation Mode}
\newacronym{smf}{SMF}{Session Management Function}
\newacronym{smo}{SMO}{Service Management and Orchestration}
\newacronym{sms}{SMS}{Short Message Service}
\newacronym{smsgmsc}{SMS-GMSC}{\gls{sms}-Gateway}
\newacronym{snr}{SNR}{Signal-to-Noise-Ratio}
\newacronym{son}{SON}{Self-Organizing Network}
\newacronym{sptcp}{SPTCP}{Single Path TCP}
\newacronym{srb}{SRB}{Service Radio Bearer}
\newacronym{srn}{SRN}{Standard Radio Node}
\newacronym{srs}{SRS}{Sounding Reference Signal}
\newacronym{ss}{SS}{Synchronization Signal}
\newacronym{sss}{SSS}{Secondary Synchronization Signal}
\newacronym{st}{ST}{Spanning Tree}
\newacronym{svc}{SVC}{Scalable Video Coding}
\newacronym{tb}{TB}{Transport Block}
\newacronym{tcp}{TCP}{Transmission Control Protocol}
\newacronym{tdd}{TDD}{Time Division Duplexing}
\newacronym{tdm}{TDM}{Time Division Multiplexing}
\newacronym{tdma}{TDMA}{Time Division Multiple Access}
\newacronym{tfl}{TfL}{Transport for London}
\newacronym{tfrc}{TFRC}{TCP-Friendly Rate Control}
\newacronym{tft}{TFT}{Traffic Flow Template}
\newacronym{tgen}{TGEN}{Traffic Generator}
\newacronym{tip}{TIP}{Telecom Infra Project}
\newacronym{tm}{TM}{Transparent Mode}
\newacronym{to}{TO}{Telco Operator}
\newacronym{toa}{ToA}{Time of Arrival}
\newacronym{tr}{TR}{Technical Report}
\newacronym{trp}{TRP}{Transmitter Receiver Pair}
\newacronym{ts}{TS}{Technical Specification}
\newacronym{tti}{TTI}{Transmission Time Interval}
\newacronym{ttt}{TTT}{Time-to-Trigger}
\newacronym{tx}{TX}{Transmitter}
\newacronym{uas}{UAS}{Unmanned Aerial System}
\newacronym{uav}{UAV}{Unmanned Aerial Vehicle}
\newacronym{udm}{UDM}{Unified Data Management}
\newacronym{udp}{UDP}{User Datagram Protocol}
\newacronym{udr}{UDR}{Unified Data Repository}
\newacronym{ue}{UE}{User Equipment}
\newacronym{uhd}{UHD}{\gls{usrp} Hardware Driver}
\newacronym{ul}{UL}{Uplink}
\newacronym{um}{UM}{Unacknowledged Mode}
\newacronym{uml}{UML}{Unified Modeling Language}
\newacronym{upa}{UPA}{Uniform Planar Array}
\newacronym{upf}{UPF}{User Plane Function}
\newacronym{urllc}{URLLC}{Ultra Reliable and Low Latency Communication}
\newacronym{usa}{U.S.}{United States}
\newacronym{usim}{USIM}{Universal Subscriber Identity Module}
\newacronym{usrp}{USRP}{Universal Software Radio Peripheral}
\newacronym{utc}{UTC}{Urban Traffic Control}
\newacronym{vim}{VIM}{Virtualization Infrastructure Manager}
\newacronym{vm}{VM}{Virtual Machine}
\newacronym{vnf}{VNF}{Virtual Network Function}
\newacronym{volte}{VoLTE}{Voice over \gls{lte}}
\newacronym{voltha}{VOLTHA}{Virtual OLT HArdware Abstraction}
\newacronym{vr}{VR}{Virtual Reality}
\newacronym{vran}{vRAN}{Virtualized \gls{ran}}
\newacronym{vss}{VSS}{Video Streaming Server}
\newacronym{wbf}{WBF}{Wired Bias Function}
\newacronym{wf}{WF}{Wired-first}
\newacronym{wi}{WI}{Wireless InSite}
\newacronym{wlan}{WLAN}{Wireless Local Area Network}
\newacronym{pnf}{PNF}{Physical Network Function}
\newacronym{drl}{DRL}{Deep Reinforcement Learning}
\newacronym{mtc}{MTC}{Machine-type Communications}
\newacronym{v2x}{V2X}{Vehicle-to-everything}
\newacronym{cast}{CaST}{Channel emulation scenario generator and Sounder Toolchain}
\newacronym{gui}{GUI}{Graphical User Interface}
\newacronym{ups}{UPS}{Uninterruptible Power Supply}
\newacronym{ota}{OTA}{Over-the-Air}
\newacronym{hitl}{HITL}{hardware-in-the-loop}
\newacronym{soc}{SoC}{System-on-Chip}
\newacronym{ecdf}{eCDF}{Empirical Cumulative Distribution Function}
\newacronym{cnn}{CNN}{Convolutional Neural Network}
\newacronym{nn}{NN}{Neural Network}
\newacronym{mpc}{MPC}{Multi Path Component}
\newacronym{otc}{OTC}{Over-the-Cable}